\documentstyle[12pt,psfig,here]{article}
\newfont{\sansb}{cmssbx10}
\newfont{\sans}{cmss10}

\begin{document}

\begin{flushright}
{\bf Preprint ISS - 8 - 1997 (December, 1997)}\\
{\bf Institute of Space Sciences, Bucharest, Romania}
\end{flushright}

\begin{center}
{\Large \bf FRAGMENTATION AND MULTIFRAGMENTATION OF 
10.6A GeV GOLD NUCLEI}\\

{\bf EMU01-Collaboration} \\
\end{center}

M. I. Adamovich$^{13}$, M. M. Aggarwal$^{4}$, Y. A. Alexandrov$^{13}$,
R. Amirikas$^{17}$, N. P. Andreeva$^1$,
S. K. Badyal$^{8}$, A. M. Bakich$^{17}$, E. S. Basova$^{18}$, 
K. B. Bhalla$^7$, A. Bhasin$^8$
% ...
\\

{\small {\it 1 High Energy Physics Institute, Almaty, Kazakstan \\
2 Institute of High Energy Physics, Academia Sinica, 
Beijing, China \\3
Institute of Space Sciences, Bucharest, Romania \\4 Department of Physics,
Panjab University, Chandigarh, India \\5 Department of Physics, Hunan
Education Institute, Changsha, Hunan, China \\6 Laboratory of High Energies,
Joint Institute for Nuclear Research (JINR), Dubna, Russia \\7 Department of
Physics, University of Rajasthan, Jaipur, India \\8 Department of Physics,
University of Jammu, Jammu, India\\9 Department of Nuclear Physics,
University of P. J. Safarik, Kosice, Slovakia \\10 Department of Physics,
Shanxi Normal University, Linfen, Shanxi, China \\11 Department of Physics,
University of Lund, Lund, Sweden \\12 F.B. Physik, Philipps University,
Marburg, Germany \\13 P.N. Lebedev Institute of Physics, Moscow, Russia \\14
V.G. Khlopin Radium Institute, St. Petersburg, Russia \\15 Department of
Physics, University of Washington, Seattle, Washington, U.S.A. \\16
Department of Physics, Kon-kuk University, Seoul, Korea \\17 School of
Physics, University of Sydney, Sydney, Australia \\18 Lab. of Relativistic
Nuclear Physics, Institute of Nuclear Physics, Tashkent, Uzbekistan \\19
Lab. of High Energies, Physical-Technical Institute, 
Tashkent, Uzbekistan \\20 Institute of Particle Physics, 
Hua-Zhong Normal University, Wuhan, Hubei,
China \\21 Institute of
Experimental Physics, Slovac Academy of Science, Kosice, Slovakia.}}

\newpage

{\em Abstract} : We present the results of a study performed on the
interactions of 10.6A GeV gold nuclei in nuclear emulsions. 
In a minimum bias sample of
1311 interactions, 5260 helium nuclei and 2622 heavy fragments were 
observed as Au projectile fragments.
The experimental data are analyzed with particular emphasis of 
target separation interactions in emulsions and study of critical
exponents.
Multiplicity distributions of the fast-moving 
projectile fragments are investigated.
Charged fragment moments, conditional moments as well as two and three -
body asymmetries of the fast moving projectile 
particles are determined  in
terms of the total charge remaining bound in the multiply charged projectile
fragments. Some differences in the average yields of helium nuclei and
heavier fragments are observed, which may be attributed to a target effect.
However, two and three-body asymmetries and conditional moments indicate
that the breakup mechanism of the projectile seems to be independent of
target mass.
We looked for evidence of 
critical point observable in finite nuclei by study the resulting 
charged fragments distributions.  
We have obtained  the values for the critical exponents 
 $\gamma$, $\beta$ and $\tau$ and  compare our 
results  with those at lower energy experiment
(1.0A GeV data). The values suggest that a phase transition like 
behavior, is observed .

\section{Introduction}

The experiment to be described here examines the breakup of relativistic
gold nuclei when they interact with the target nuclei in nuclear emulsions.
The fragments produced are readily identified in the emulsions. Specific
attention in this paper is directed towards the multiply charged fragments
that are produced. Some reports on the main characteristics of the
interactions have been published 
previously \cite{emu01_1}-\cite{emu01_10}.

Depending upon the target - projectile combination and the incoming
projectile energy, the excited piece of nuclear matter decays predominantly
by the emission of nucleons, deuterons, tritons, helium nuclei and charged
particles with $3 \leq Z \leq 30$ commonly known as intermediate - mass
fragments (IMF's) and fragments of very heavy charge $Z \geq 31$. To
understand the dynamics involving the formation of helium, IMF's and other
multi-fragments in its final state, numerous experiments have been performed
at low, intermediate and high energies, in both p-nucleus and
nucleus-nucleus reactions \cite{gori61}-\cite{ritt94}.

The experiment to be described here examines the breakup of relativistic
gold nuclei when they interact with the target nuclei in nuclear
emulsions. The relative yields of the different types of fragment and the
relationships between them are measures of processes that occur during the
breakup of the excited nuclear remnants. We will study some relationships
and compare with those observed in others experiments. In this paper we
shall present a systematic study on the target and projectile fragmentation
of the $^{197} Au$ - induced emulsion interactions at $10.6A\,\,GeV$ from the
BNL Alternating Gradient Synchrotron (AGS).

Competing models suggest different decay mechanism and experiments have yet
to discriminate between several theoretical scenarios which range from the
sequential decay of the compound nucleus \cite{fried90,char88} to
statistical nuclear models \cite{gros91,bond85} and percolation models
\cite{bauer88,stau79}.

It has been stressed out by Campi \cite{camp86} that the moments of the charge
distributions provide a test of the suggestion that multifragmentation can
be described in terms of percolation theory. If there is some critical
behavior in the breakup of the nuclei, such as a liquid - gas phase
transition, then some events should have values of normalized moments much
larger than the average. For the low energy gold interactions \cite{wad85}
there was a wide range of values and strong and approximately linear
correlations between the different normalized moments, although it was not
possible to conclude that then was a phase transition. Also EOS collaboration 
\cite{gilk94} have reported some of their results from the analysis of 1.0A
GeV gold nuclei fragmenting in a carbon target. In their analysis of 9716
interactions they used the methods developed for determining percolation
critical exponents to extract the values of 
specific exponents for nuclear matter from
the moments of the fragment charge distribution.

However, one of the main problem encountered in 
interpreting the results from nuclear emulsion experiments 
is the non-homogeneous composition of 
 the emulsion, which contains
both light (H, C, N, O) and heavy (Ag, Br) target nuclei.
Critical remarks about using minimum biased samples for studing
critical behavior have been expressed \cite{jakob90}. It has 
been argued that in emulsion experiments the mixture between 
emission sources, both with respect to origin and size, should be 
a severe shortcoming for collisions from few hundreds of MeV/nucleon 
up to several GeV/nucleon \cite{jakob90}.

Therefore one of the main  objective of this
paper is to present a detailed  
analysis of specific  measured quantities for multifragmentation phenomenon  
for a larger sample of these interactions of
gold nuclei, with special emphasis on the inclusive interactions with
separated light and heavy target nuclei.

 Recently EMU01-Collaboration \cite{emu01_10} using a statistical
analysis based on event by event charge distributions, showed that a 
population of sub-critical, critical and supercritical events was 
observed among peripheral collisions, but the study has been 
limitted only to critical exponent $\gamma$, which in some papers
\cite{gilk94} is claimed to show little  sensitivity 
to the system under investigation .  

KLMM Collaboration \cite{cher95} has also
looked for evidence of phase changes in the description of
multifragmentation at 10.6A GeV, but their results are significantly
inconsistent with those reported at lower energies,
suggesting that percolation theory
becomes a less satisfactory representation of the breakup for these high
energy interactions than it was at lower energies.

Therefore it will be of
great interest to repeat such analysis at the same energy and study
the values for all critical exponents, using the same methods.

\section{Experimental Details}

Several stacks of electron sensitive, NIKFI BR-2 emulsions,
of dimension $10$ $X$ $10$ $X$ 
$0.06$ $cm^3$ have been exposed to a 
$10.6A$ $GeV$ $\;^{197}Au$ beam at BNL synchrotron (experiment E863).
The sensitivity of emulsion was about 30 grains per unit 
length of $100$ $\mu m$ for singly charged particles with minimal ionization.

Primary interactions were found by along - the track double scanning : fast
in the forward direction and slow in the backward direction. Fast scanning
was made with a velocity excluding any discrimination of events in the
number of heavily ionizing tracks, slow scanning was carried out to find
events, if any, with little changed and unbiased projectile nucleus. This
analysis immediately resulted in a determination of the mean free path (mfp)
for interactions. The measured mean free path $\lambda = 4.99 \pm 0.16$ $cm$
agreed well with KLMM - collaboration result $\lambda = 4.7 \pm 0.2$ $cm$
\cite{cher95} and an approximation of measured cross-sections on various nuclei
and targets $\lambda = 4.6$ $cm$  \cite{binn87}.

In each event the polar angles $\theta $ and azimuthal angles $\varphi$
were measured. Depending on ionization, all tracks emitted from the
interaction vertices were classified according to the commonly accepted
emulsion experiment terminology :

\begin{quote}
1)Shower, or {\bf s} - particles - singly charged particles with a velocity 
$\beta \geq 0.7$ ; 
they, mainly, consist of produced particles and singly charged
fragments.

2)Grey, or {\bf g} - tracks, whose ionization (the number of grains per unit
length ) correspond to protons with momentum 
$0.2 \leq p \leq 1 \; GeV/c$ ; they
consist, mainly of protons knocked - out from the target nucleus during the
collision with a few percent admixture of $\pi$ mesons with 
momentum $60 \leq p \leq 170 \; MeV/c.$

3)Black, or {\bf b} - tracks - mostly, protons with 
$p \leq 0.2 \; GeV/c$ and
multiply charged target fragments. They have a range of $ < 3 \; mm.$

4) Fragments of projectile nucleus - particles with $Z \geq 2.$ Fragments
with $Z = 2 \;$are identified by visual inspection of tracks, their ionization
is constant and equal to $g/g_0 \approx 4$ , $g_0$ - ionization due to
relativistic particle (minimal ionization).

To determine charge of fragments with $Z \geq 3$, the density was measured
of $\delta $ electrons on length no less than $ 10 \; mm$ ; the beam track and
track with $Z=2$ were chosen as tracks for calibrating. 
The accuracy of
charge measurement was for $Z < 10,\;\pm 1\;$for 
$10 \leq Z < 28,\;\pm 2\;$for $28 \leq Z < 40\;$and 
$\pm 3\;$for $Z \geq 40\;$ in units of charge.

Grey and black tracks amount the group of heavily ionizing 
tracks $N_h = N_g + N_b.$

In each event, the number of produced particles, $\pi $ 
mesons ($N_{\pi} )\;$ have been also determined.

A number of 1311 inelastic interactions 
were obtained after excluding from
the ensemble events those of electromagnetic origin and pure elastic
scattering. Others details on experiment 
especially on charge fragment measurements and errors,
 have been recently published \cite{emu01_10}.
\end{quote}

\section{Projectile breakup}

\subsection{Average multiplicity of fast - moving particles.}

The difference between the projectile and the target spect-ator fragments is
easy to make. Projectile fragments and shower particles are very energetic and
they are distributed in a forward narrow cone. 
The angle of this cone is $< \theta_{0} >$ = 16.4 mrad
and it can be found from the pseudorapidity
distribution of the shower particles.

The target fragments are observed as highly ionizing particles,
isotropically distributed. They can be {\bf black} particles which are
essentially evaporation fragments from the target, with $R < 3\; mm$,
or {\bf gray} particles which are knock-out protons or 
slow mesons with $R > 3\; mm$ .

The breakup of the projectile can be characterized 
in terms of the numbers, $N_F$ - and charges , 
$Z_F$, of the fragments with $Z \geq 3$, that are
emitted ; the number $N_\alpha$, of alpha particles emitted and the numbers 
$N_{prot}$, of protons released and available to interact, where from
charge balance we can write :

\begin{equation}
N_{prot} = Z_{Au}- \sum_{Z_F \geq 3} N_F*Z_F-2 * N_{\alpha}  
\label{eq:nprot}
\end{equation}

We obtain the bound charge by $Z_{bound}\;$,by :

\begin{equation}
Z_{bound} = \sum_{Z \geq 2} n(Z)*Z
\label{eq:zbound}
\end{equation}

where $n(Z)$ is the multiplicity of the projectile fragments 
with $Z \geq 2$.

In Table I , we display results of the present investigation of the average
multiplicities $< N_{prot} >,\,< N_\alpha >,\,$
 and $< N_F >$ of the fragments
with $Z = 1,\,Z = 2,\,$ and $\,\,Z \geq 3,$ respectively,
for the $^{197} Au$ ions.
\newpage
Also a comparison with the results of Cherry et al. \cite{cher95} and those
of Singh and Jain \cite{jain96} at 10.6A GeV are given. 
A good agreement with both experiments is obtained 
for $< N_{prot} >$ and $< N_\alpha >$,
but a slight difference is remarked for mean number of projectile
fragments with $Z \geq 3$.

\begin{table}[H]
\caption{The average multiplicity of projectile fragments 
with $Z = 1,\,Z = 2,\, and\,\,Z \geq 3$ 
emitted in the $^{197}Au$ - induced reactions in emulsions at 10.6A GeV.}
\vskip 0.8cm
\begin{center}
\begin{tabular}{||c||c|c|c|c|c||}
\hline\hline
Beam & Energy & $< N_{prot} >$ & $< N_\alpha >$ & $< N_F >$ & Ref. \\ 
& (AGeV) &  &  &  &  \\ 
\hline\hline
$^{197} Au$ & $10.6$ & $28.48 \pm 0.81$ & $4.63 \pm 0.13$ & 
$2.01 \pm 0.06$ & \cite{jain96} \\
\hline 
$^{197} Au$ & $10.6$ &  & $4.53 \pm 0.13$ & $1.91 \pm 0.06$ &
\cite{cher95} \\ 
\hline
$^{197} Au$ & $10.6$ & $28.44 \pm 0.64$ & $4.51 \pm 0.08$ & 
$2.37 \pm 0.03$ & This work \\
\hline\hline
\end{tabular}
\label{tab:mult1}
\end{center}
\end{table}

The numbers of fragments emitted from individual interactions can be
compared by looking at the distributions of the numbers of events with a
given number of fragments divided by the total number of events, the
fractional yield. These fractional yields for fragments, those with 
$Z \geq 3$ (Fig. 1a), for alpha particles (Fig. 1b), 
for heaviest fragments (Fig. 1c)
and for numbers of released protons (Fig. 1d) are compared 
in Figure 1 with the
published results of KLMM collaboration \cite{cher95,wad93}. 
It appears that
these yields are nearly identical, although some differences appear
especially for heaviest fragment and number of proton released. 

The experimental spectra for number of released protons (Fig. 2a) and
number of produced pions (Fig. 2b) as well as their correlation with number
of mesons (Fig. 2c) and the number 
of heavily ionizing tracks $N_h$ (Fig. 2d) are given in Figure 2. As
would be expected, there is a strong correlation between the number of
proton released and the number of pions produced - there is a rather well
defined upper bound to the number of pions 
$N_\pi $ for a given number of proton released $N_{prot}$. 
Fig. 2d shows that
interactions with light target nuclei, those with $N_h \leq 7$, never
results in a copious pion production. For $N_h > 8\;$, where the target must
have been a AgBr nucleus, $N_\pi \;$ appears to be almost 
independent of $N_h$.

It is interesting to note that our data set for $^{197} Au$ 
do not give any evidence of the occurrence 
of binary fission in the charge range of $35 \leq Z \leq 45$ 
\cite{emu01_9}, although a significant enhancement of the fission
events was observed at lower energies with the $^{197} Au$ projectile, as
reported in Ref. \cite{wad85}.

\subsection{Correlation between $N_\alpha $ and $N_F$}

Broad characteristics of the projectile fragmentation can be explored
through the correlation between the multiplicity 
of helium particles $N_\alpha \;$ and heavier fragments $N_F$ . 
Such a correlation is given in
Table 2 for the $^{197} Au$ data . From Table 2 one can see that the number
of pure central interactions, in which the projectile has disintegrated
completely into singly charged fragments with no emission of heavier
fragments, is very small ($\approx 1\,\,\%$). 
More than $20\,\, \%$ of the events
have $N_\alpha = 0$ and $N_{F} = 1$. Only $2.5\,\, \%$ of the total number of
interactions is found with $N_F \geq 5$ and 
$0 \leq N_\alpha \leq 10$. 
On the other hand, only $2.1\,\, \%$ of the total umber of
collisions is observed with $N_F < 5 \;$ and $N_\alpha > 10$.
 No event is
observed with $N_F \geq 8\;$ and $N_\alpha > 10$.
 The maximum number of
fragments with $Z \geq 3$ in our sample is $N_F = 8$ 
and that of fragments with 
$Z \geq 2$ is $N_\alpha + N_F = 21$.

\begin{table}[H]
\caption{Characteristics of projectile fragmentation 
of the $^{197} Au$
beam through a correlation between the number of helium particles,
$N_\alpha$ and the number of heavy fragments, $N_F$ with $Z \geq 3.$}
\vskip 0.8cm
\begin{center}
\begin{tabular}{||r||r|r|r|r|r|r|r|r|r||}
\hline
\hline
      & $N_F = 0$ & 1 & 2 & 3 & 4 & 5 & 6 & 7 & 8 \\ 
\hline\hline
$N_\alpha $ = 0 & 16 & 165 & 231 & 17 & 1 & 0 & 0 & 0 & 0 \\ 
\hline
1 & 21 & 90 & 18 & 6 & 1 & 1 & 1 & 0 & 0 \\ 
\hline
2 & 14 & 67 & 30 & 14 & 4 & 0 & 0 & 0 & 0 \\ 
\hline
3 & 18 & 70 & 36 & 19 & 13 & 3 & 21 & 0 & 0 \\ 
\hline
4 & 9 & 50 & 50 & 34 & 10 & 8 & 4 & 0 & 1 \\ 
\hline
5 & 3 & 33 & 40 & 33 & 19 & 2 & 3 & 0 & 0 \\ 
\hline
6 & 3 & 24 & 31 & 21 & 14 & 5 & 2 & 0 & 0 \\ 
\hline
7 & 1 & 20 & 28 & 21 & 12 & 6 & 3 & 0 & 0 \\ 
\hline
8 & 5 & 16 & 22 & 14 & 8 & 4 & 0 & 1 & 0 \\ 
\hline
9 & 3 & 10 & 16 & 7 & 12 & 6 & 1 & 0 & 0 \\ 
\hline
10 & 2 & 9 & 13 & 7 & 5 & 1 & 0 & 0 & 0 \\ 
\hline
11 & 0 & 4 & 4 & 3 & 4 & 2 & 0 & 0 & 0 \\
 \hline
12 & 0 & 0 & 1 & 2 & 1 & 0 & 0 & 0 & 0 \\
 \hline
13 & 0 & 1 & 1 & 3 & 0 & 0 & 0 & 0 & 0 \\ 
\hline
14 & 0 & 2 & 0 & 0 & 0 & 0 & 0 & 0 & 0 \\ 
\hline
15 & 0 & 0 & 1 & 1 & 0 & 0 & 0 & 0 & 0 \\ 
\hline
16 & 1 & 1 & 0 & 0 & 0 & 0 & 0 & 0 & 0\\
\hline
\hline
\end{tabular}
\end{center}
\label{cor:nfna}
\end{table}

Specific correlations for mean number $< N_F + N_\alpha >$ with number of
singly charged fragments $(N_P)$ are shown in Fig. 3a. We can remark
an expected linear correlation for $N_P \leq 10$ and a plateau for
$N_P > 10$. Correlations with number of shower particles $N_S$
  are given in Fig. 3b,
( we note that $N_S$ does not include singly charged fragments $N_P$)
  indicate a clear change in  slope for 
 $N_S > 70 $, which is due to separation of central and peripheral 
 collisions. A {\em rise} and {\em fall} for correlations of   
 mean number $< N_F + N_\alpha >$ with number of released proton,
  which are complementary to correlations with $Z_{bound}$
 (see also Fig. 5a,d) are depicted  in Fig. 3c. 
 A more complex correlations are obtained by plotting  
   $< N_F + N_\alpha >$ as function of number of black and grey 
 tracks ($N_h$) in Fig. 3d, in which we can see a different slopes 
 for interactions with light ($N_h < 7$) and heavy nuclei 
 ($N_h \geq 8$.

\subsection{Target separation}

Nuclear emulsion are a composite medium composed of AgBr, CNO and H.
Certainly, they are also other nuclei in emulsions, but their concentration
are too small to be taken into account. It was a difficult task to separate
interactions on different classes of targets and it is impossible to find
certain separation criteria that give no admixture between those classes
although there are many correlations between the measurable parameters that
give informations regarding the target nuclei.

The separation technique we used was based on the analysis of specific
correlations between target break-up and particle production. Depending upon
the target break-up, we separate the sample of 1311 interactions into three
classes :

a) $N_h = 0, 1$ ; this class includes all Au + H interactions but also
interactions with other targets.

b) $2 \leq N_h \leq 7$, containing the rest of Au + CNO interactions not
included at a), but also some Au + AgBr.

c) $N_h > 7$, including only interactions with AgBr.

An additional relation that give some target separation for 
 a) events is based on 
the distribution of the number of shower particles 
with $\theta > \theta_0 $. 
Events with no black or gray tracks include most of the H
interactions, but also the most central events with CNO and AgBr. The
distribution of the shower particles for these events 
indicates  the limit of no
more than about 40 relativistic particles with $\theta > \theta _0$ emitted
from Au + CNO interactions. The separation between H and CNO peaks give an
admixture of less than $10\,\, \%$ for this class of events.

For $N_h = 1$ we analyzed separately elastic interactions from a kinematic
approach. Knowing the emission angles for the black track and the recoiled
gold nucleus we could compute the range-momentum ratio for most of the
elastic events. This ratio is highly dependent on the target size and it's
very useful in order to separate elastic interactions with CNO from those
with AgBr. For elastic interactions with hydrogen nuclei, the recoil proton
appear most often as a gray track. The number of shower
particles for inelastic interactions with $N_h = 1$ indicates that 
AgBr events are for $ N_{s^{\prime }} > 40$.

Class b) events were separated by plotting the number of shower particles
against $N_h$. The separation between CNO from AgBr populations is given by :

\begin{equation}
N_{s^{\prime}} < 175 - 14.5 \cdot N_h
\label{eq:nsprim}
\end{equation}

where, $N_{s^{\prime}} = N_s + N_P$.

The reason we took the limit $N_h > 7$ for class c) events was that 8 target
tracks for an oxygen nucleus correspond to a total charge break-up, 
reasonable to assume that at least
one of the released target protons become a relativistic particle.

We plotted the distributions of the number of $\alpha $ particles and PF's
for the interactions of gold nuclei with HCNO and AgBr 
targets in Figure 4.
We can see in Fig. 4a that the emission of $\alpha $ particles is enhanced
for interactions with heavy targets, the distribution being shifted to the
right. However the distributions of heavy fragments in Fig. 4b show the same
behavior for the two samples. 

The correlations between the mean number of emitted alpha's and fragments
and the total bound charge are depicted in Fig. 4c and Fig. 4d. We can see
that for peripheral events corresponding to $Z_{bound} > 50$ the emission of
alpha's and fragments have the same behavior both for light and heavy
emulsion targets. However $\alpha $ particle emission is suppressed almost
down to zero when centrality is increasing, especially for HCNO
interactions. The difference between the two samples for PF's is not so
obvious. Nevertheless we can notice some decrease of the mean number of
fragments for HCNO compared to AgBr in semi-central events.

\section{Specific Correlations for Multifragmentation}

\subsection{Correlation between some observable and $Z_{bound}$.}

The parameter $Z_{bound},\;$ which was defined by 
Hubele et al., \cite{ogi91}
is related to the size of the projectile spectator and the energy deposited
in a given collision can also be explored through $Z_{bound}$.
 $Z_{bound}$ is always larger than or equal to $Z_{max}$.
 
The heaviest fragment charge in each collision, $Z_{max}$, provides useful
information on the exit channel of that collision. Fig. 5a shows a scatter
plot of the correlation between $Z_{max}\;$and $Z_{bound}$ for individual
events of the $^{197} Au$ data. We remark that most of the data points are
situated below the diagonal, that means that $Z_{bound}$ is always larger
than or equal to $Z_{max}$ . Also in the peripheral collisions, the
largest fragment contains most of the total bound charge and in such
reactions $Z_{max}$ may be identified as the heavy residue of the beam
nucleus after evaporation. For the central events, we observe 
that $Z_{max}$ becomes a smaller fraction of 
$Z_{bound}$ . This plot also confirms that symmetric fission is a very
rare kind of events.

In Fig. 5b we give the variation of mean value of 
$ < Z_{max} > / Z_{beam}$ as
a function of $Z_{bound}$ for the $^{197} Au$ data. A sharp rise of this
ratio can be seen from Fig. 5b for $Z_{bound}$ $\;\approx\; 40-79.$ 
For $Z_{bound}\; \approx\; 2-40$ the rise is rather slow. This plot give some
information into the degree of breakup of the projectile nucleus.

In Fig. 5c we plot the average multiplicity distributions of alpha particles 
$< N_\alpha >\;$ and in Fig. 5d the mean number of intermediate - mass
fragments (IMF's) with $3\; \leq\; Z\; \leq\; 30$ as a function of 
$Z_{bound}\;$for $^{197} Au$ projectiles. Our experimental data confirm, 
the so called ''rise and fall'' of the multi-fragment emission .

\subsection{Charged fragment asymmetries}

In this chapter, we investigate the charge asymmetry between 
the  largest fragments ($Z_{max}$) and the second largest $Z_{FM2}$
fragment in an event, using the two body relative asymmetry $ R_{AS}$,
defined as \cite{ogi91} :

\begin{equation}
R_{AS} = \frac{Z_{max} - Z_{FM2}}{Z_{max} + Z_{FM2}}  
\label{eq:ras}
\end{equation}

Only fragments with $ Z \geq 2$ are included.

Also $R_{AS1}$, the asymmetry  between the second 
largest $Z_{FM2}$ and third largest $Z_{FM3}$ in an event is obtained by :

\begin{equation}
R_{AS1} = \frac{Z_{FM2} - Z_{FM3}}{Z_{FM2} + Z_{FM3}}  
\label{eq:ras1}
\end{equation}

We can also investigate the breakup process in a more qualitative manner
through another parameter known as three body asymmetry \cite{ogi91} as :

\begin{equation}
R_{AS2} = \frac{\sqrt{(DFA)^2 + (DFB)^2 + (DFC)^2}}{\sqrt{6} < Z > }  
\label{eq:ras2}
\end{equation}

where :

\begin{center}
$DFA = Z_{max} - < Z >$

$DFB = Z_{FM2} - < Z >$

$DFC = Z_{FM3} - < Z >$
\end{center} 

with :

\begin{equation}
< Z > = \frac{1}{3} (Z_{max} + Z_{FM2} + Z_{FM3})  
\label{eq:zmed}
\end{equation}

The parameter $R_{AS2}$ has a maximum near unity when there is heavy residue
of the projectile spectator and has a zero value when the projectile
fragments of equal size are emitted in the collisions.

For the analysis of two body asymmetries given by 
Eq. (\ref{eq:ras}) and Eq. (\ref{eq:ras1}) , 
we have selected the events with at least 
three fragments of $Z \geq 2$. 

In Fig. 6a and Fig. 6b, we plot a correlation between mean value 
of ratio $< R_{AS} >$
and $< R_{AS1} >$ respectively with $Z_{bound}$ for $^{197} Au$ data.
The ratio $< R_{AS} >$ decrease monotonically 
from its maximum value $\approx 0.9$ to
almost zero as one approaches from extremely peripheral toward more violent
collisions. The ratio $< R_{AS1} >$ rises linearly with $Z_{bound}$ 
for $Z_{bound}\; \leq\; 40$ and decreases for 
$40$ $\leq\; Z_{bound}\; \leq\; 79$ for these
class of events selected $(N_F + N_\alpha\; \geq\; 3)$ .
In Fig. 6c we represent ratio $< R_{AS} >$ for IMF's and 
we see that the ratio remain almost constant 
for $20\; \leq\; Z_{bound}\; \leq\; 60$. 
For $Z_{bound}\; > \; 60$ a clear leading fragments effects appear.
In Fig. 6d we plot a graph of a correlation between mean value
$< R_{AS2} >$ and $Z_{bound}$.
 The parameter $< R_{AS2} >$ rises almost 
linearly with enhancing value of $Z_{bound}$.

In order to observe if there is any difference in the behavior of the
projectile break-up mechanisms for interactions with different targets in
emulsion, we analyzed two and three-body asymmetries between the heaviest
emitted fragments.

In this analysis we used together the separation criteria for H and CNO
interactions in order to eliminate any possible admixture between these
groups. Thus we built two samples, the first containing interactions with
light (HCNO) nuclei and the second AgBr events.

In Figure 7 we plot the same parameters for interactions with light target
(HCNO) (open circles) and heavy targets (Ag,Br) (full circles) as
described in section 3.3.

We plotted the correlations between the mean two and three-body asymmetry
ratios for interactions with light and heavy nuclei and the total bound
charge. We can see in Fig. 7a that the asymmetry ratio between the first and
second heavy fragments increases monotonically both for light and heavy
samples from most central interactions to peripheral one 
($Z_{bound}\; \simeq\; 79 $). The second and third fragment 
asymmetry ratio (Fig . 7b) seems to be
constant for semi-central and peripheral interactions, decreasing down to 0
for $Z_{bound} < 40$ when increasing the centrality.

The asymmetry ratio for IMF's (see Fig. 7c) is constant 
in a wide region of $Z_{bound}$, peaking near the 
value of 70 which indicate a higher asymmetry
between the IMF's in peripheral interactions. The three-body asymmetry ratio
plotted in Fig. 7d show an approximately linear dependence with $Z_{bound}$
but also there is no significant difference between the two samples.

These observations point out that multifragmentation of gold nuclei at this
energy does not depend on the target nucleus. Thus it doesn't matter that
nuclear emulsions contain an admixture of targets from the
multifragmentation mechanisms point of view.

\subsection{Moments of the Charge Distribution}

Multifragmentation has been considered to be one of the most important
aspects of heavy - ion collisions since it has been speculated that the
decay of a highly excited nuclear system might carry information about the
equation of state and the liquid-gas phase transition of low density nuclear
matter. 
The similarity between statistical multifragmentation models and
percolation theory has been stressed in ref. \cite{bauer88,stau79}
The relevance of percolation ideas in nuclear fragmentation 
can be investigated
better by examining cross relations between various moments of the fragment
size distribution. We will show in this paper that experimental data have
strong similarities with the predictions of percolation models.

Following Campi \cite{camp86} suggestion we investigate the moments of the
charge distribution of the $n_Z$ projectile fragments (PF's) using an
event-by-event based analysis. For each event, we determine the 
multiplicity of charged fragments, $m_{PF}$, and the number of charged
fragments, $n_Z$, of nuclear charge Z. We then consider the $i'th$ 
moments of this distribution :

\begin{equation}
M_{i} = \sum_{Z} Z^{i}*n_{Z} 
\label{eq:mi}
\end{equation}

where the sum is extended over all the fragments except 
the biggest cluster (fragment) 
which is being considered as the percolating cluster 
\cite{stauf92}. Physically $Z_{max}$ corresponds to the bulk liquid in
an infinite system. 

The zero order moment
is obtaining by taking $i\,=\, 0$ in Eq. (\ref{eq:mi}) :

\begin{equation}
M_{0}= \sum_{Z} n_{Z}
\label{eq:m0} 
\end{equation}

It has also been suggested \cite{camp86} that the conditional moment, 
$\gamma_2$, which is combined from  the moments $M_2$, $M_0$ and $M_1$
as :

\begin{equation}
\gamma_2 = \frac{M_2 M_0}{M_{^1}^2}  
\label{eq:gamma2}
\end{equation}

 give more selective information.

In Figure 8 the mean values of $M_0$ (Fig. 8a), $M_1$ (Fig. 8b), and $M_2$
(Fig. 8d) , averaged over small range of $Z_{bound}$ for the nonzero values
of the moments, are shown as function of $Z_{bound}$ . In Fig. 8c we
represent a variation of mean values of $<\gamma _2>$ for the events with at
least three charged projectile fragments with $Z\geq 2.$ 
The value of $<\gamma _2>$ increases rather slowly in the 
range of $2\; \leq\; Z_{bound}\; < 50$.
The maximum value of $< \gamma_2 >\,\, \approx\,\, 1.35$ at 
$Z_{bound}\; \approx\; 50$ and
then decreases for $Z_{bound}\; > \;50$. Some fluctuations 
in $< \gamma_2 >$ can be
seen in this region for the $^{197} Au$ data. 
For the values of $Z_{bound}\; \leq\; 20$, $< \gamma_2 >\,=\, 1$ 
(events in which the projectile fragments
of the same size are emitted). For the infinite nuclear system, the scaling
theory of critical phenomena predicts that at the critical 
point $\gamma_2$
diverges at a rate that depends on the critical indices of the phase
transition \cite{camp86}. For a finite nuclear system, $\gamma_2$ is
predicted to have a smooth behavior \cite{camp86}. From Fig. 9c we remark a
smooth increase of $< \gamma_2 >$ up to 
$Z_{bound}\; \approx\; 60$ for the $^{197} Au$ data. 
However some fluctuations appear in our $^{197} Au$ sample .
Similar results were analyzed for $^{197} Au$ data at 
10.6A GeV and $^{208} Pb$
data at 160A GeV \cite{jain96}.
The peak observed is due to finite size of the
nuclear systems under investigation.

In Figure 9 we study  the dependence of the same 
quantities $M_0$ (Fig. 9a), $M_1$ (Fig. 9b), 
$M_2$ (Fig. 9d) and $\gamma_2$ (Fig. 9c) 
function of $Z_{bound}$ for different interactions 
on light targets (open circles) and
heavy targets (full circles). No significant differences between
the two target groups are observed.

\section{Critical Exponents in Multifragmentation}

\subsection{Leading fragments}

Fig. 10a, exhibits an area plot of the correlation between
the charge of the heaviest fragment 
$Z_{max}$ and the first order moment $M_1$ 
( or remaining bound charge $Z_r$) . This shows that there are many
interactions where $M_1\, >\, Z_{max}$ and that, even when there is a well
defined leading fragment (a fragment that carries more than half of the
bound charge $Z_{bound}$) - it is usually accompanied by appreciable other
bound charge.
 
\begin{equation}
Z_{bound} = Z_{max} + M_1 = Z_{max} + Z_r
\label{eq:zbound1}  
\end{equation}

The plot of $Z_{max}$ versus $M_1$, already presented in Fig. 10a shows a
clear, but complex correlation, which suggest two distinct
populations.
This suggestion is enhanced by separating the interactions 
into those with ( $M_1\, < \,Z_{max}$) and without 
($M_1\, >\, Z_{max}$) leading fragment. 

If the mean values of $ < Z_{max} >$ are
determined over small intervals of $M_1$, for these two classes they 
will appear well separated (Fig. 10c). 

A similar comparison can be drawn from the correlation between the next
heaviest fragment $Z_{FM2}$ and $Z_{max}$ in Fig. 10b. 
The mathematical restrictions 
are indicated by the dasheded lines on this plot, but it is clear
that the allowed space is not uniformly populated. In particular the fission
region near the apex is almost empty (see also reference \cite{emu01_9}), in
contrast to the situation seen at lower energies (\cite{wad85}).

Considering the ratio $R_1$ for all fragments, where :

\begin{equation}
R_1 = \frac{Z_{max }}{M_1}  
\label{eq:R1}
\end{equation}

Then there is a leading fragment, as defined above , 
when $ R_1 > 1$ . We see
from Fig. 10d that ratio $R_1$ is a strong function of $Z_{bound}$.
Thus, while there is a well defined leading 
fragment when $Z_{bound}$ is greater
than $60$, it is much less well defined when $Z_{bound}$ falls to
$40$.

\subsection{Critical Exponents}

 The observation of a power law behavior for the size distribution 
of the fragments has triggered a number of studies that have looked
for evidence of {\em critical behavior} \cite{gilk94}, \cite{cher95},
\cite{jain96}, \cite{emu01_10}. These analysis consider nuclear
multifragmentation as one example of a {\em critical phenomenon} 
and attempts are made to extract from the data the related critical 
exponents.

The EOS Collaboration \cite{gilk94} have reported some of their results
from the analysis of 1.0A GeV gold nuclei fragmenting in a carbon target
using the methods developed for determining percolation critical exponents
to extract critical exponents  for nuclear matter from the moments of the
fragment charge distribution \cite{elliot94} :

We assume that the multiplicity of fragments, 
$m = N_F + N_A + N_{prot}$ is a linear
measure of the distance from the critical point as suggested by Campi
\cite{camp86}. Here $N_F$ - is multiplicity of 
fragments ($Z \, \ge\,3$, $N_A$ is multiplicity of alpha particles 
($Z\,=\,2$ and $N_{prot}$ is the number of released protons 
($N_{prot}\, =\, Z_{beam} - Z_{bound}$ with  $Z_{beam}\, =\, 79$). 

The region in $m$ below the assumed critical 
multiplicity $m_c$ is designated as
the {\em liquid phase }and that above $m_c$ as the {\em gas phase}. 
It is assumed that in the liquid phase the 
heaviest fragment $Z_{max}$ is
omitted in forming the moments, but is not omitted when in the gas
phase \cite{stauf92}.
 
Also our analysis tacitly assumes that all of the 
projectile-related charges are associated with 
multifragmentation and include in the analysis of the moments  
number of released protons, $N_{prot}$ as fragments,
modifying Eq. (\ref{eq:mi}) and replacing $M_{i}$ by $M_i^{*}$.

The critical exponents $\gamma$, $\beta$, and $\tau$ for large 
systems are given by the following equations in 
terms of the multiplicity difference, 
$\zeta = m - m_c$ by :

\begin{equation}
M_2^{*}\;\sim \;\;|\zeta ^{-\gamma }| 
\label{eq:m2star}  
\end{equation}

\begin{equation}
Z_{max}\;\;\sim \;|\zeta |^\beta 
\label{eq:zatbeta}
\end{equation}

\begin{equation}
n_Z\;\;\;\sim \;Z^{-\tau} 
\label{eq:nzattau}
\end{equation}

These exponents $\gamma ,\,\,\beta $ and 
$\,\,\tau $ are related by the equation \cite{stan71} :

\begin{equation}
 \tau = 2 + \frac {\beta}{\beta\, +\, \gamma } 
\label{eq:sc_ab}
\end{equation}

We start our analysis by examining the variation of the mean values for
second moments $< M_2^{*} >$ as a function of multiplicity $m$ which is
depicted in Fig. 11a . A relatively abrupt change in this distribution is
apparent for $m_c$ around $26$, suggesting that there could be a phase
change at this critical value of $m_c$ . This values is similar to those
reported by others experiments \cite{gilk94}, \cite{cher95}. 

Fig. 11b shows a log-log plot of mean values of second moments
$< M_2^{*} >$  with $\zeta$ for the assumed liquid and gas 
phases, setting $m_c = 26$ ( where log stand for natural logaritms).
The clear separation between the two phases arises 
from the inclusion of $Z_{max}$
in the determination of $< M_2^{*} >$ in the gas phase. 
In Fig. 11c and Fig. 11d the values are represented in a scatter plot.

If we examine over the entire available range of $\zeta$ , neither phase
shows the power law behavior predicted by Eq.
\ref{eq:m2star}  . However, if rather
narrow regions of $|\zeta |$ are selected, 
 $5\,\, \leq\,\, |\zeta | \,\,\leq\,\, 20$, then a good fit to such a power
low can be obtained for the gas phase , 
with $\gamma_{gas} = 0.86 \pm 0.05$
and a reduced $\chi ^2$ of $2.94$. A fair fit can also be obtained for the
liquid phase, with 
$\gamma_{liquid} = 0.83 \pm 0.14$, reduced $\chi ^2$ of $1.44$. 
 The results are represented in Fig. 12a.

We note that the values are relatively sensitive to the range of $|\zeta |$
used, as we expected, since finite size distorsions dominate as 
$|\zeta|\,\, \rightarrow\,\,0$, and signature of critical behavior 
vanish for large $|\zeta|$, corresponding to mean field regime. 
 Adding two more values to those shown in Fig. 12a changes 
$\gamma $ to $\gamma_{liquid} = 0.69 \pm 0.11$ 
with $\chi ^2$ of $1.56$ and $\gamma_{gas} = 0.73 \pm 0.04$ 
with $\chi ^2$ of $4.3$.

The most important results is that the values for $\gamma_{liquid}$ are
close to the values of $\gamma_{gas}$ , which implies that the conditions
for a phase change have been satisfied. 
No better match can be found for $m_c = 30$.
In practice to estimate the uncertainties, we varied the fitting
region by changing the upper and lower multiplicity limits. The overall
estimated uncertainties are $14.5\,\, \%$.

We continue the analysis determining the exponent $\beta $ 
from Eq. \ref{eq:zatbeta}
considering the liquid phase, where $Z_{max}$ is well defined. Fig. 12b
shows $\log (< Z_{max} >)$ as a function of $|\zeta |$ . The value obtained
for $\beta = 0.25 \pm 0.02$ with a reduced $\chi ^2 = 1.56$ is in a good
agreement with that of $0.29 \pm 0.02$ reported at 
1.0A GeV \cite{gilk94} (see also table \ref{tab:cexp}).
The value obtained for an exponent is sensitive to the range of values
chosen for $|\zeta |$ .

The critical exponent $ \tau $ in Eq. (\ref{eq:nzattau}) can be 
determined from the slope of $\log (< M_3^{*} >)$ versus 
$\log (< M_2^{*} >)$ (see Fig. 12c) using only the
gas phase \cite{camp86}. 
Fig. 12d shows a power law fit with $\tau = 2.23 \pm 0.05$ and a reduced 
$\chi ^2$ of $1.51$, for $3\,\, \leq \,\,|\zeta |\,\, \leq\,\, 36 $. 

Our value is practically the same with the value predicted in infinite
percolation models \cite{stau79} or the values reported for the 1.0A GeV 
\cite{gilk94} $2.26$ or the value of $2.23$ calculated from our measured
values of $\beta $ and $\gamma_{gas}$ using Eq. (\ref{eq:sc_ab}).

Reducing the 
range of $|\zeta| $ values used for this fit does not make any 
significant change in the deduced values of $\tau $. 
 We note that the critical exponent $\tau$ is close to 2.2 for many
three-dimensional systems and thus does not permit a determination 
of the universality class of phase transition.

The exponent values are summarized in table \ref{tab:cexp}, 
which also list the values from  
lower energies experiment \cite{gilk94}, percolation 
\cite{stauf92}, liquid-gas values \cite{stan71} and the mean field
limit of the liquid-gas system \cite{call85}.

\newpage

\begin{table}[h]
\caption{Critical multiplicity and exponents for Au projectile
  fragmentation and other three-dimensional systems} 
\vskip 0.3cm 
\centering
 \begin{tabular}{|c|c|c|c|c|} \hline\hline
 {\bf Quantity} &$\bf{m_C}$ & 
$\bf{\gamma}$ & $\bf{\beta}$ & $\bf{\tau}$ \\
\hline\hline
Our Exper. & 26 & $0.86 \pm 0.05$ & $0.25 \pm 0.02$ & $2.23 \pm 0.05$\\
\hline\hline
EOS Exper.\cite{gilk94} & $26 \pm 1$ & $1.40 \pm 0.1$ & 
$0.29 \pm 0.02$ & $2.14 \pm 0.06$  \\
\hline\hline
Percolation  \cite{stauf92}  &  & 1.8 & 0.41 & 2.18  \\
\hline\hline
Liquid - gas &  & 1.0 & 0.50 & 2.33   \\
mean field \cite{call85}&   &   &   &    \\
\hline\hline
Liquid-gas \cite{stan71}  &  & 1.23 & 0.33& 2.21  \\
\hline\hline
\end{tabular}
\label{tab:cexp}
\end{table}

We note that the values of 
$\gamma,\,\,\beta\,\,and \,\,\tau$ obtaining using 
this method obey the scaling relation, Eq. (\ref{eq:sc_ab}). 
By varying the fitted 
region, we have obtained exponents which differ by as much as $15\,\, \%$.
We can take this to be a measure of the uncertainty in their values. 
It is significant that the values for $\beta $ and $\gamma$ are
different from either the percolation or the mean field values.
Also we can remark that $\gamma $ exponent is different from 
those of nominal fluids and depends on energy.

\section{Conclusions}

In the present work we have studied, the properties of the projectile
associated particles emitted in interactions of the $^{197} Au$ ions
accelerated at an energy of $10.6A\,$ $GeV$ obtained from the BNL AGS.

The average multiplicities of the fast - moving projectile particles such as 
$< N_\alpha >,\;< N_F >,\;< N_{prot} >$ seem to depend upon the mass of the
target. The majority of the multiply charged fragments are helium nuclei,
while the majority of those fragments with $Z \geq 3$ are light. The
multifragment emission is a dominant reaction channel 
as observed when the distributions
of $< N_\alpha > $, 
$< N_{IMF} >$ are represented as function of $Z_{bound}$.
These distributions are peaked at $Z_{bound}\; \approx\; 35-40$ 
and shows slight dependence on the target mass.

Nuclear emulsion detectors provide an excellent tool to study the global
characteristics of nucleus-nucleus interactions since they allow a
simultaneous investigation of the processes of nuclear fragmentation and
multiple particle production and allow a study of the correlations between
these processes.

Even if the emulsion detector contains different targets,
multifragmentation, when expressed
in terms of $\;Z_{bound}$ , appears to be relatively insensitive to the
nature of the target and the results can be compared with those from studies
using pure targets.

An analysis of the moments $M_0$ , $M_1$ and $M_2$ as well
as conditional moments such as $\gamma_2$ also proves that the breakup
mechanisms has no dependence on the target size and a broad peak 
in the $\gamma_2$ - $Z_{bound}\;$ relation  
shows that the nuclear systems employed in the
present investigation induced a finite size effect.

Two and three body asymmetries are explored through 
the distributions of $< R_{AS} >$,$< R_{AS1} >$ and  
$< R_{AS2} >$ as a function of $Z_{bound}$ . Within
statistical errors, the distributions shows almost similar behavior on
different target which indicates that the breakup 
mechanism has no dependence on target mass. 

 A study of multiplicities suggest that there could be a phase
change in the residual nucleus that depends on the multiplicity of the
charge fragments, in a manner similar to that predicted by theories such as
percolation that describe the process of multifragmentation. 

Our analysis
for a critical point and a phase change based on our high energy data give
results which are consistent with the analysis reported for the low energy
results. The presence of a critical point is well established from our
data. Comparison with percolative and liquid-gas systems show
remarkable similarities.
However, some essential differences on values of critical
exponents are pointed out.

 To further characterize this phenomenon, we must
determine whether the fragmenting system is thermalized and if so 
its temperature and density \cite{nupecc97},
and also whether the multiplicity is proportional to temperature. 
Some results have been 
published recently \cite{emu01_12,emu01_13}.

\section{Acknowledgements}

Financial support from the Swedish Natural Science Research Council, the
Int. Sci. Foundation and the Russian Foundation of Fundamental Research, the
German Federal Minister of Research and Technology, the University Grants
Commission ANR the Department of Science and Technology of the Government of
India, the National Natural Science Foundation of China, the Foundation of
the State Education Commission of China, the Grant Agency for Science at the
Ministry of Education of Slovak Republic and the Slovak Academy of Sciences
and the US Department of Energy, the German Academic Exchange Service
(DAAD), the Australian Research Council, Kon-Kuk University Research
Fund, the Romanian Ministry of Research and Technology and the
National Science Foundation are cordially and gratefully acknowledged.

%\epsfxsize=6.0in
%\epsfysize=6.0in
%\epsfbox{fig1.eps}

\newpage
\begin{figure}[H]
\vspace{1.cm}
\hspace{1.5cm}
\psfig{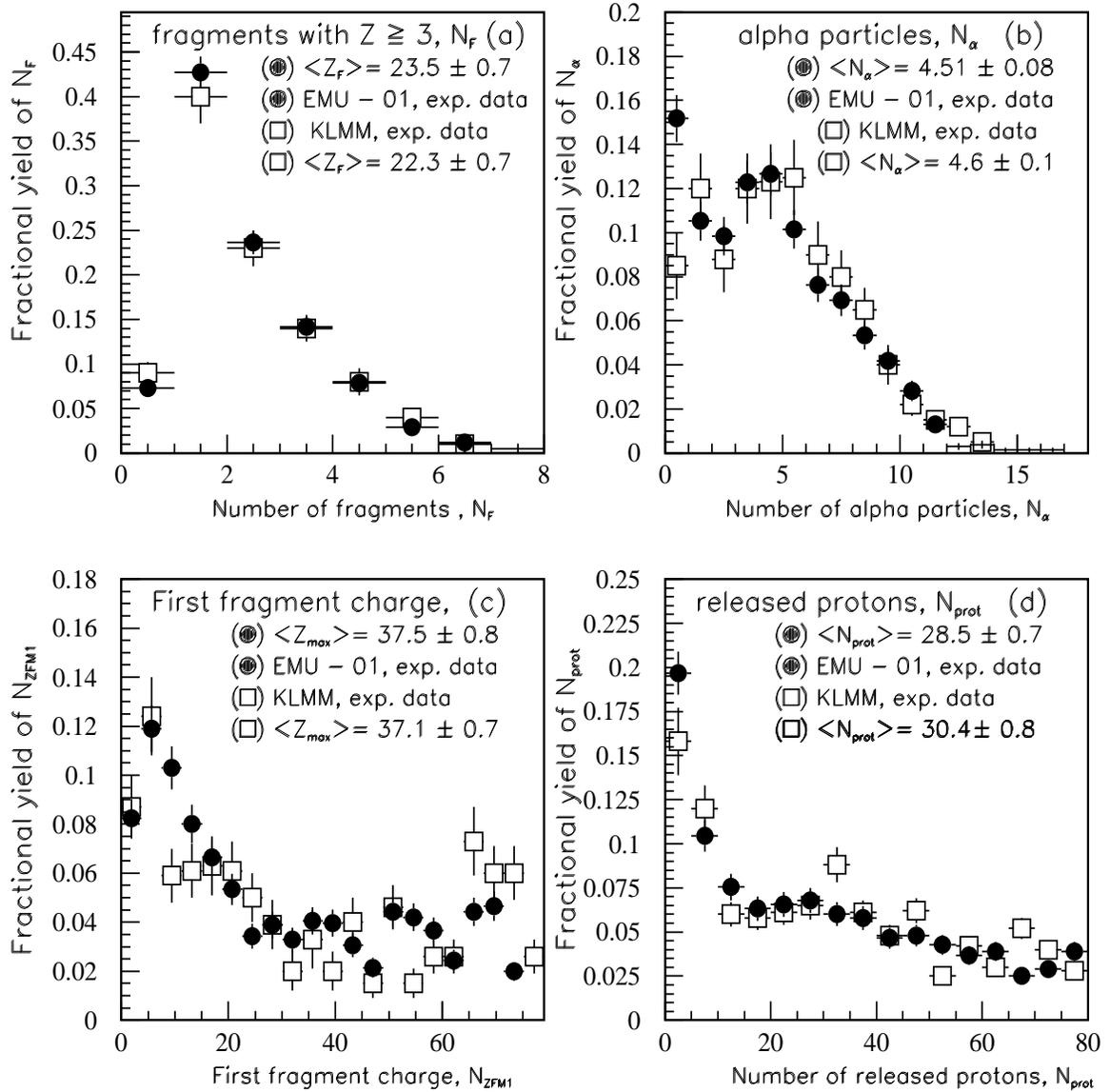}
\caption{ Part (a) - Fractional yields of fragments with $ Z \geq 3$ observed
in the two experiments ; EMU-01 data (full circles) and KLMM data
(open squares). Part (b) - Fractional yields of alpha particles in the
two experiments as a function of alpha particles emitted from an interaction.
Part (c) - Fractional yields of the heaviest 
fragments in the same experiments as
a function of the charge of the fragment. Part (d) - Fractional yields of
released protons as a function of the numbers of protons released from an
interaction.}
\end{figure}

\newpage
\begin{figure}[H]
\vspace{1.cm}
\hspace{1.5cm}
\psfig{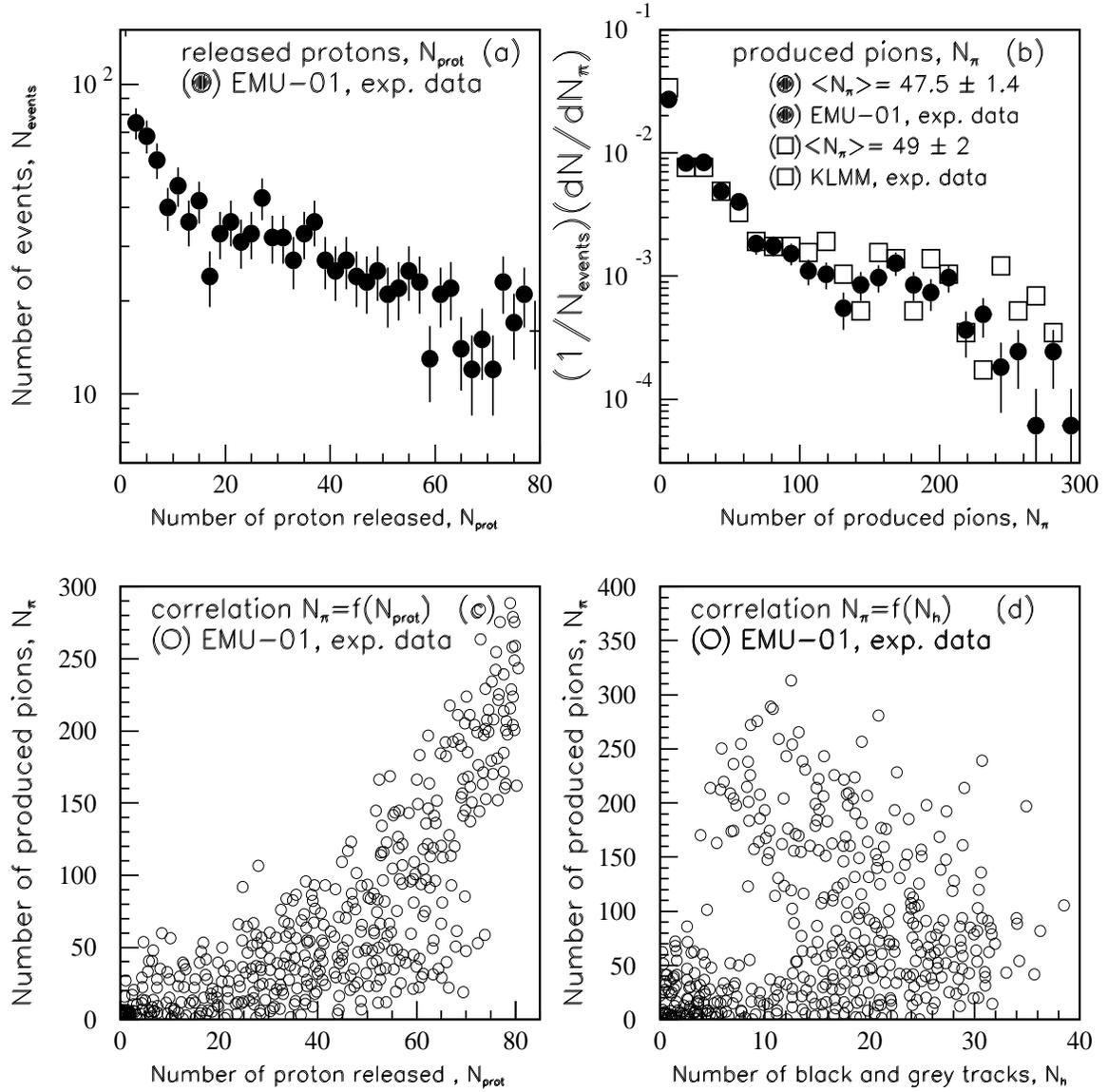}
\caption{ Part (a) - Distribution of number of proton released - full circles
EMU-01 data. Part (b) - a comparison of the number of pion produced in the
two experiments. EMU -01 data (full circles) ; KLMM data (open squares).
Part (c) - Correlation between the number 
of protons released and the number of
pion produced . Data from EMU-01 experiment. Part (d) - Number of pions
produced as a function of the number of black and grey tracks ($ N_{h}$)
emitted from the target.}
\end{figure}

\newpage
\begin{figure}[H]
\vspace{1.cm}
\hspace{1.5cm}
\psfig{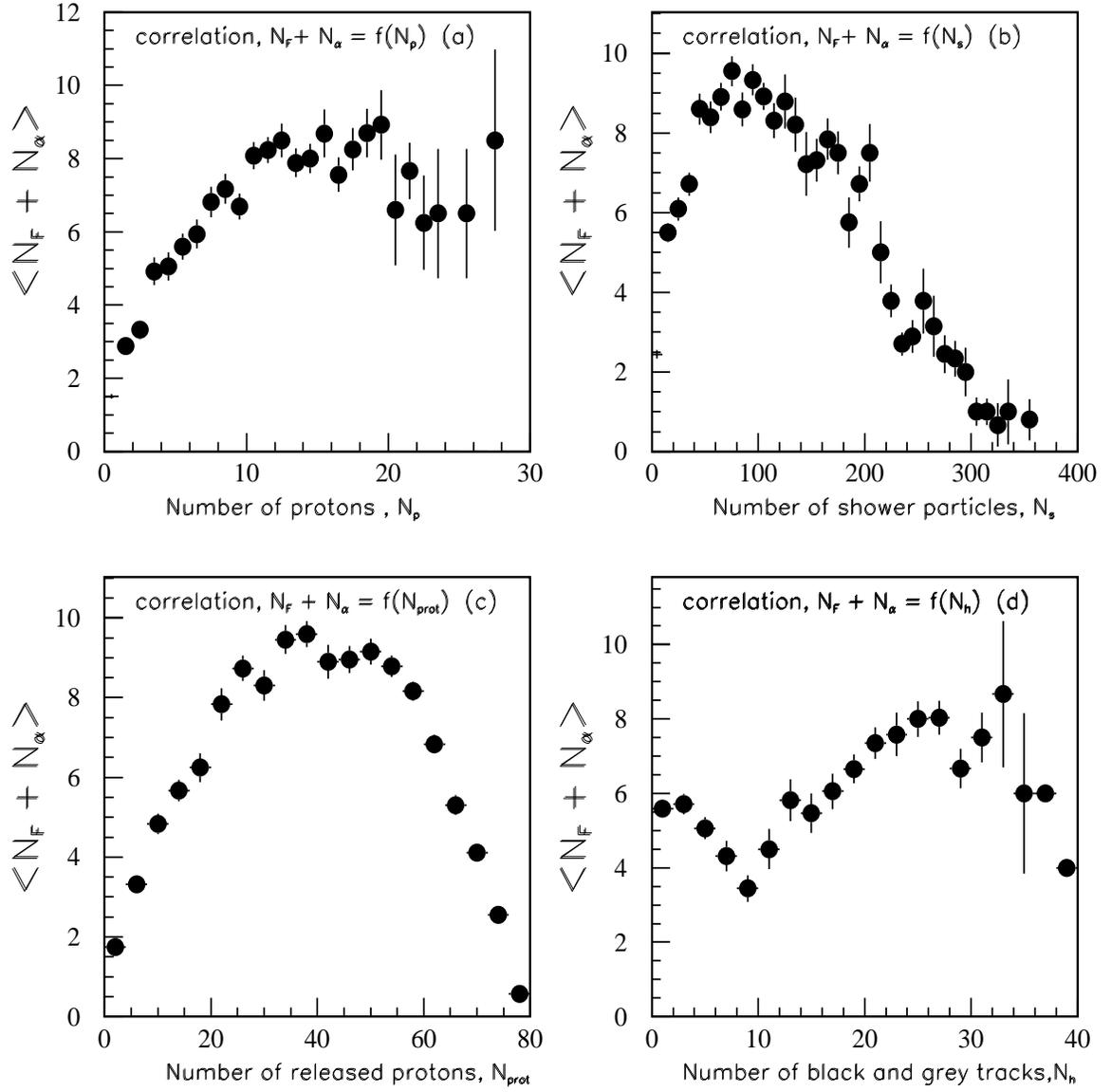}
\caption{ Correlation of mean number of alpha particles and
fragments with $ Z \geq 3$ ( $< N_{\alpha} + N_{F} >$) with number
of singly charged fragments ($N_P$)(Fig. 3a), 
with the number of shower particles ($N_S$) (Fig. 3b),
with the number of released protons ($N_{prot}$)(Fig. 3c) 
and with the number of black and grey
tracks ($N_h$)(Fig. 3d).}
\end{figure}

\newpage
\begin{figure}[H]
\vspace{1.cm}
\hspace{1.5cm}
\psfig{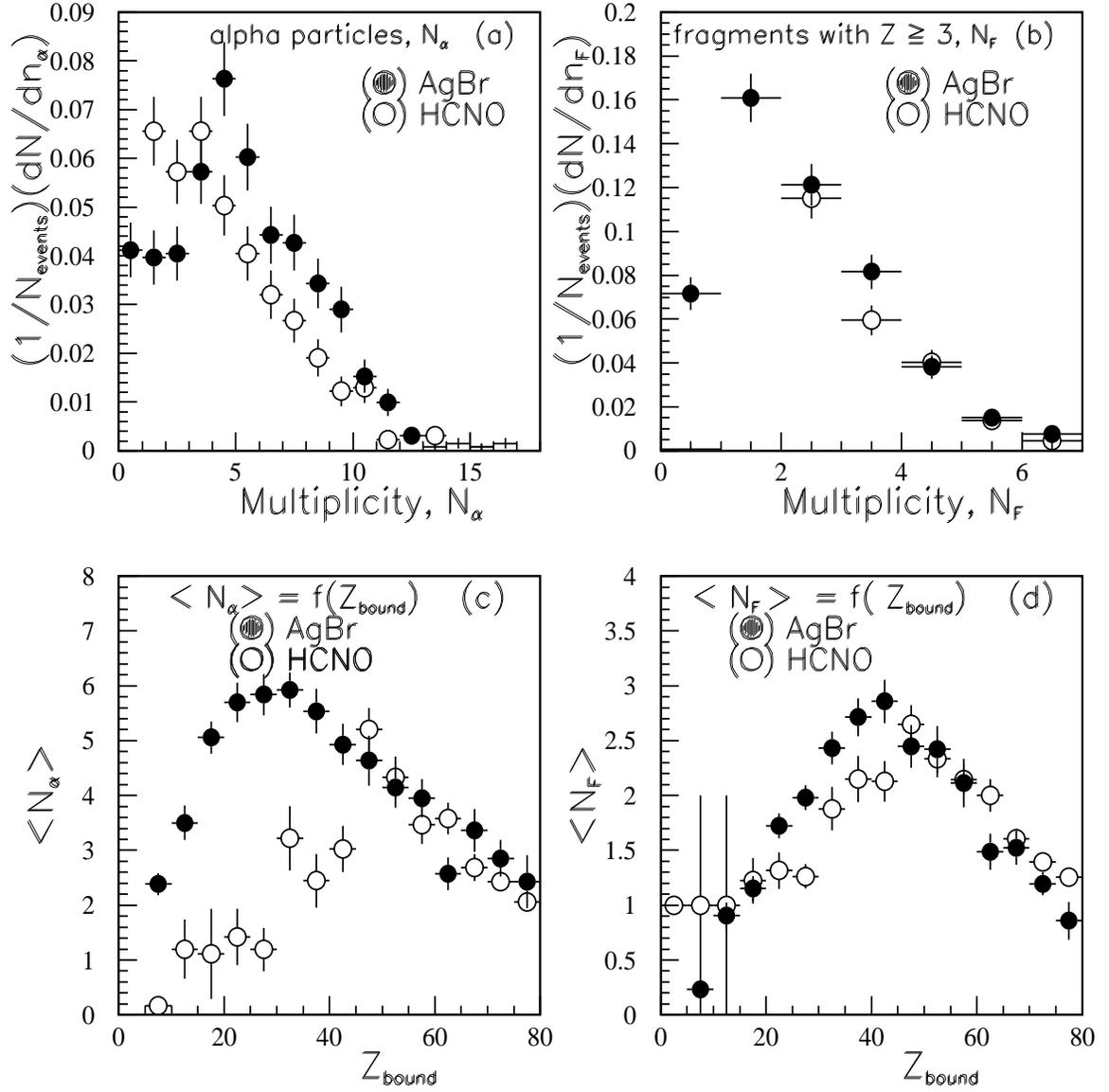}
\caption{ Part (a) - The distributions of the numbers of alpha particles
in interactions with a emulsion light (HCNO) ( open circles) and heavy
(Ag,Br) (full circles)- EMU-01 data. Part (b) - The distributions
of the numbers of fragments with $ Z \geq 3$. Part (c) - The average number
of alpha particles $ < N_{\alpha} >$ as a function of $ Z_{bound}$.
Part (d) - The average numbers of fragments ( $Z \geq 3$ ) as a function of
$ Z_{bound}$. The experimental data have the same meanings as in Part (a).}
\end{figure}
\newpage

\begin{figure}[H]
\vspace{1.cm}
\hspace{1.5cm}
\psfig{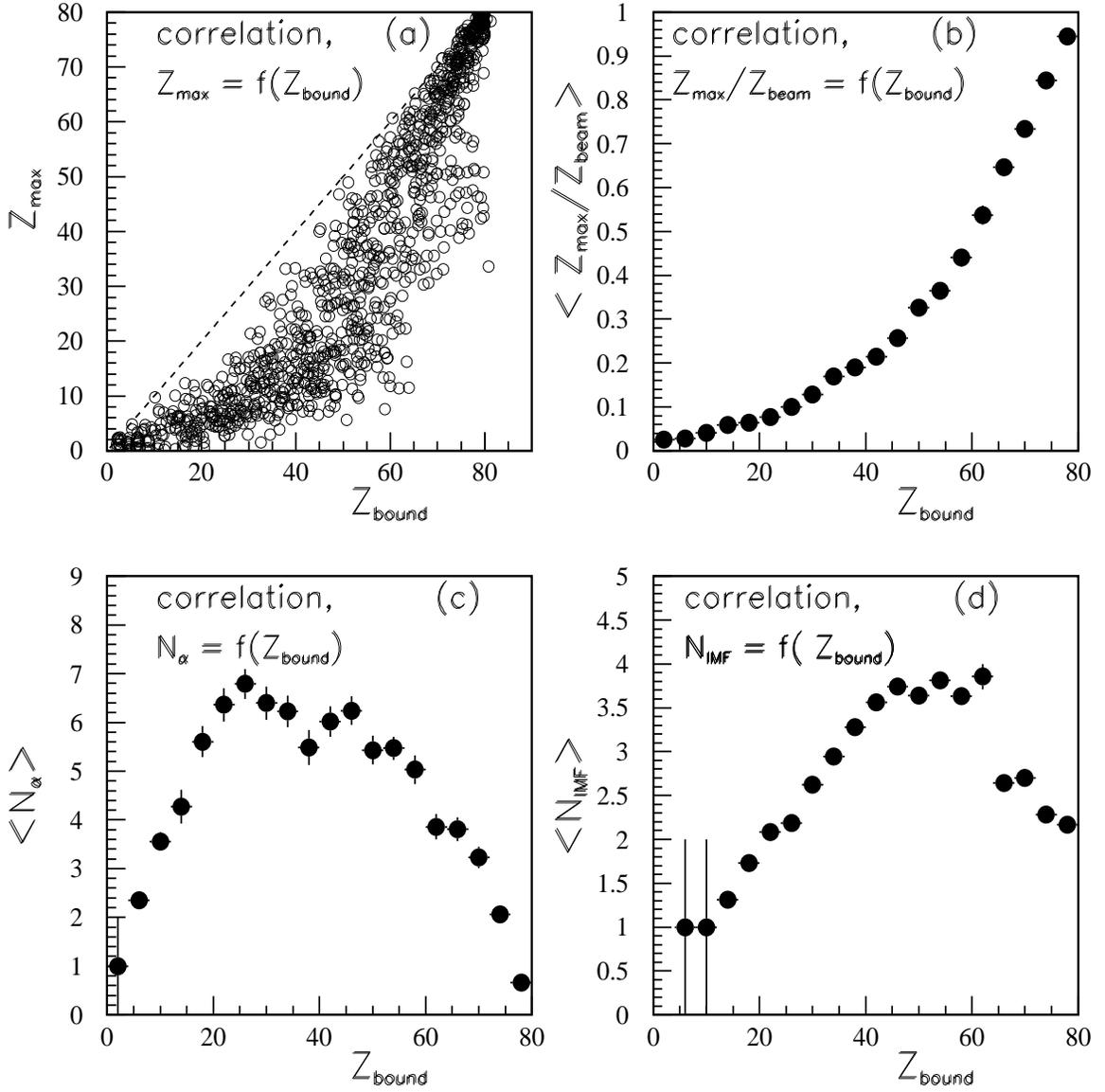}
\caption{ Part (a) - A scatter plot between the correlation of largest
charge $ Z_{max}$ and $ Z_{bound}$, in individual events . The diagonal
is shown by a dashed line. Part (b) - Mean values ration
$ < Z_{max}/ Z_{beam} > $ as a function of $ < Z_{bound} > $. Part (c) -
Mean numbers $ < N_{\alpha} >$ as a function of 
$ < Z_{bound} >$ for full sample.
Part (d) - Mean numbers of intermediate 
fragments (IMF's) $ < N_{IMF} >$ (with
 $ 3\; \leq\; Z\; \leq\; 30$) as a function of $ Z_{bound}$.}
 \end{figure}

\newpage
\begin{figure}[H]
\vspace{1.cm}
\hspace{1.5cm}
\psfig{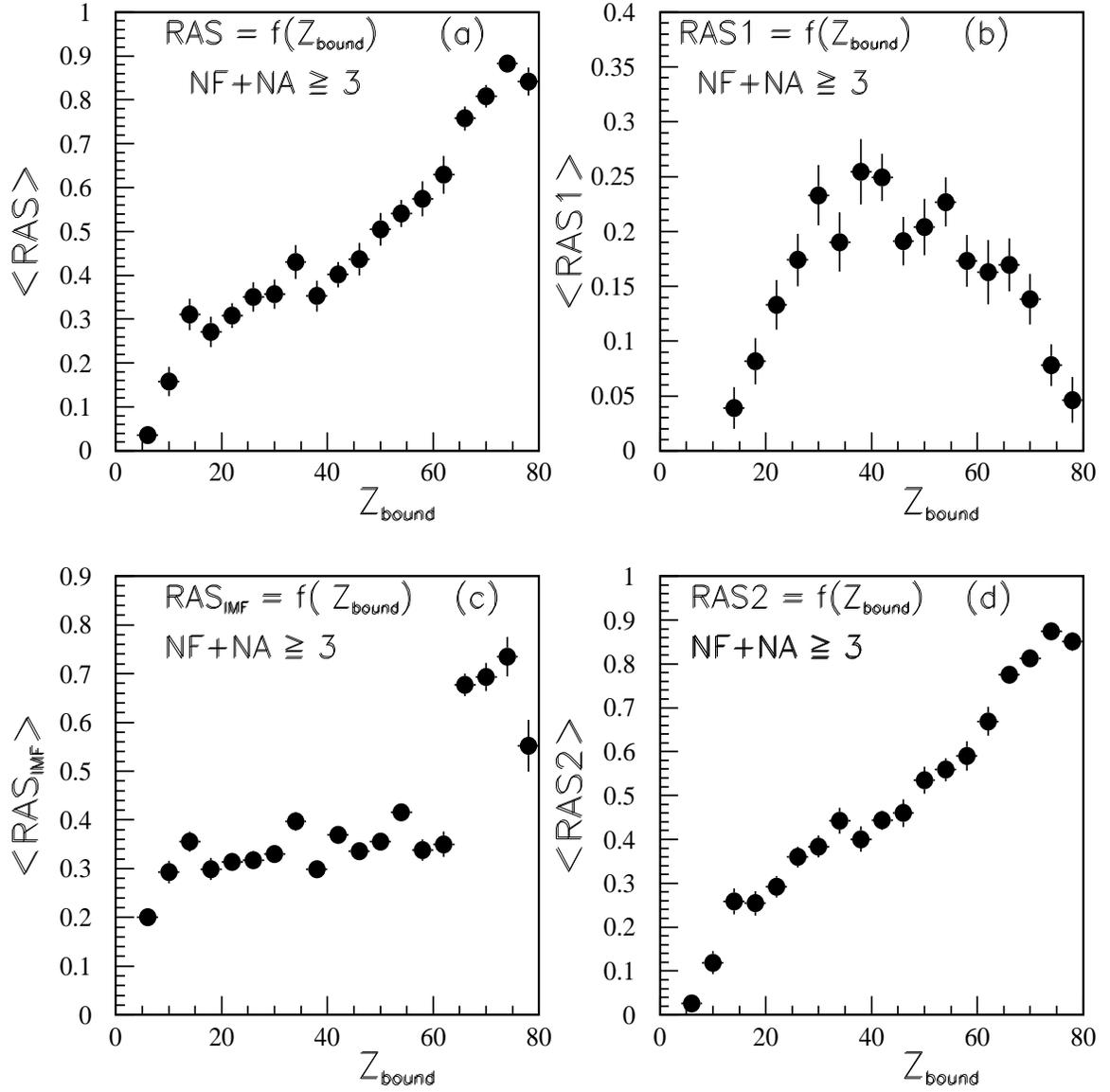}
\caption{Part (a) - Two body relative asymmetry $ < R_{AS} > $ 
versus $Z_{bound}$.
Part (b)- Two body relative asymmetry $ < R_{AS1} >$ versus $ Z_{bound}$.
Part (c) - Two body relative asymmetry for IMF's versus $ Z_{bound}$.
Part (d) - Three body asymmetry $ < R_{AS2} >$ versus $ Z_{bound}$.
For the definitions of charged particle asymmetries $< R_{AS} >$,
$ < R_{AS1} >$, $ < R_{AS2} >$ see section 4.2 }
\end{figure}

\newpage
\begin{figure}[H]
\vspace{1.cm}
\hspace{1.5cm}
\psfig{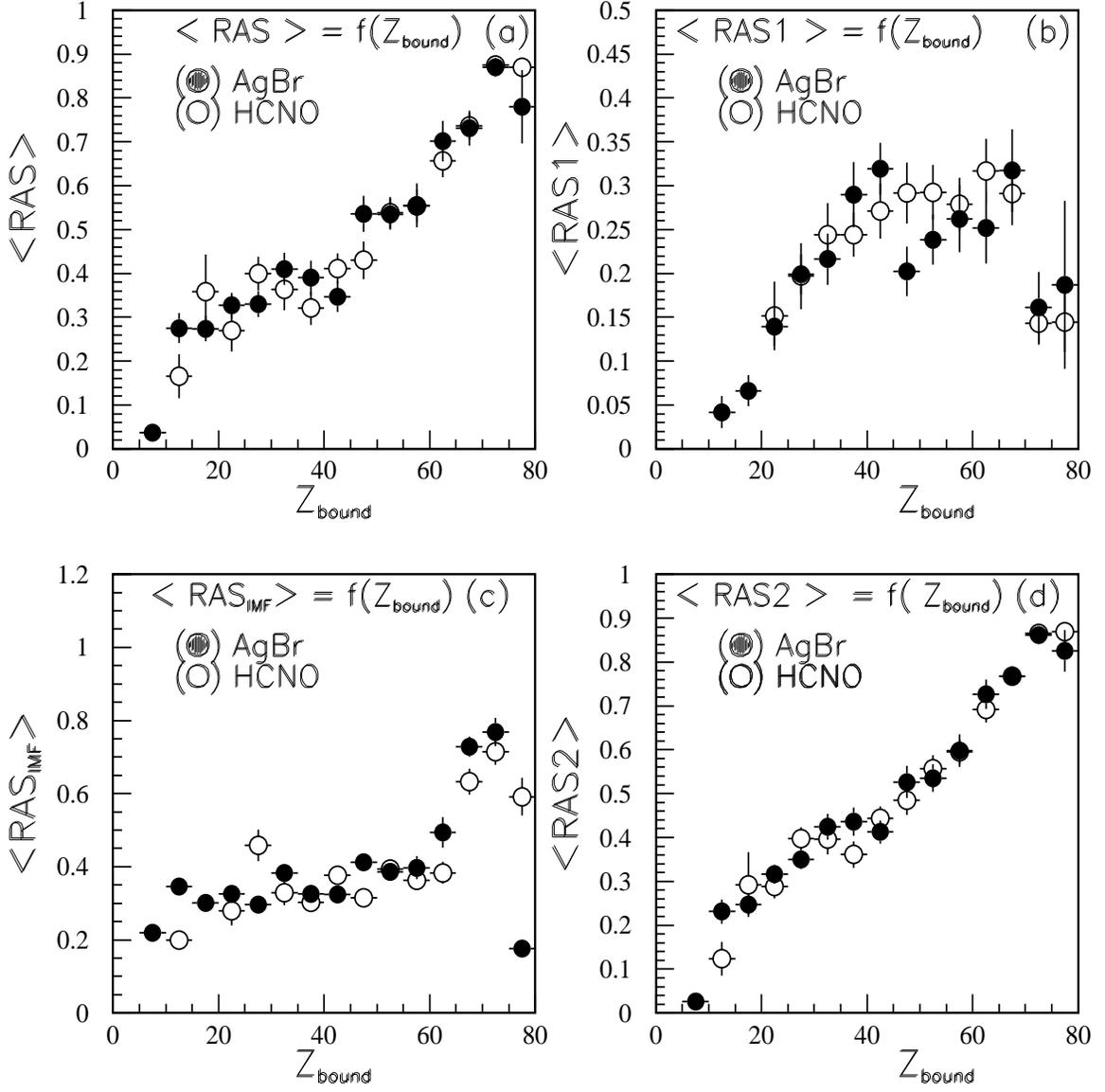}
\caption{ Part (a) - Two body relative asymmetry $< R_{AS} >$ 
versus $Z_{bound}$.
Part (b)- Two body relative asymmetry $ < R_{AS1} >$ versus $ Z_{bound}$.
Part (c) - Two body relative asymmetry for IMF's versus $ Z_{bound}$.
Part (d) - Three body asymmetry $ < R_{AS2} >$ versus $ Z_{bound}$.
Experimental points have the same meanings as in Figure 4.
For the definitions of charged particle asymmetries $ < R_{AS} >$ ,
$ < R_{AS1} > $ , $ < R_{AS2} >$ see section 4.2.}
\end{figure}

\newpage
\begin{figure}[H]
\vspace{1.cm}
\hspace{1.5cm}
\psfig{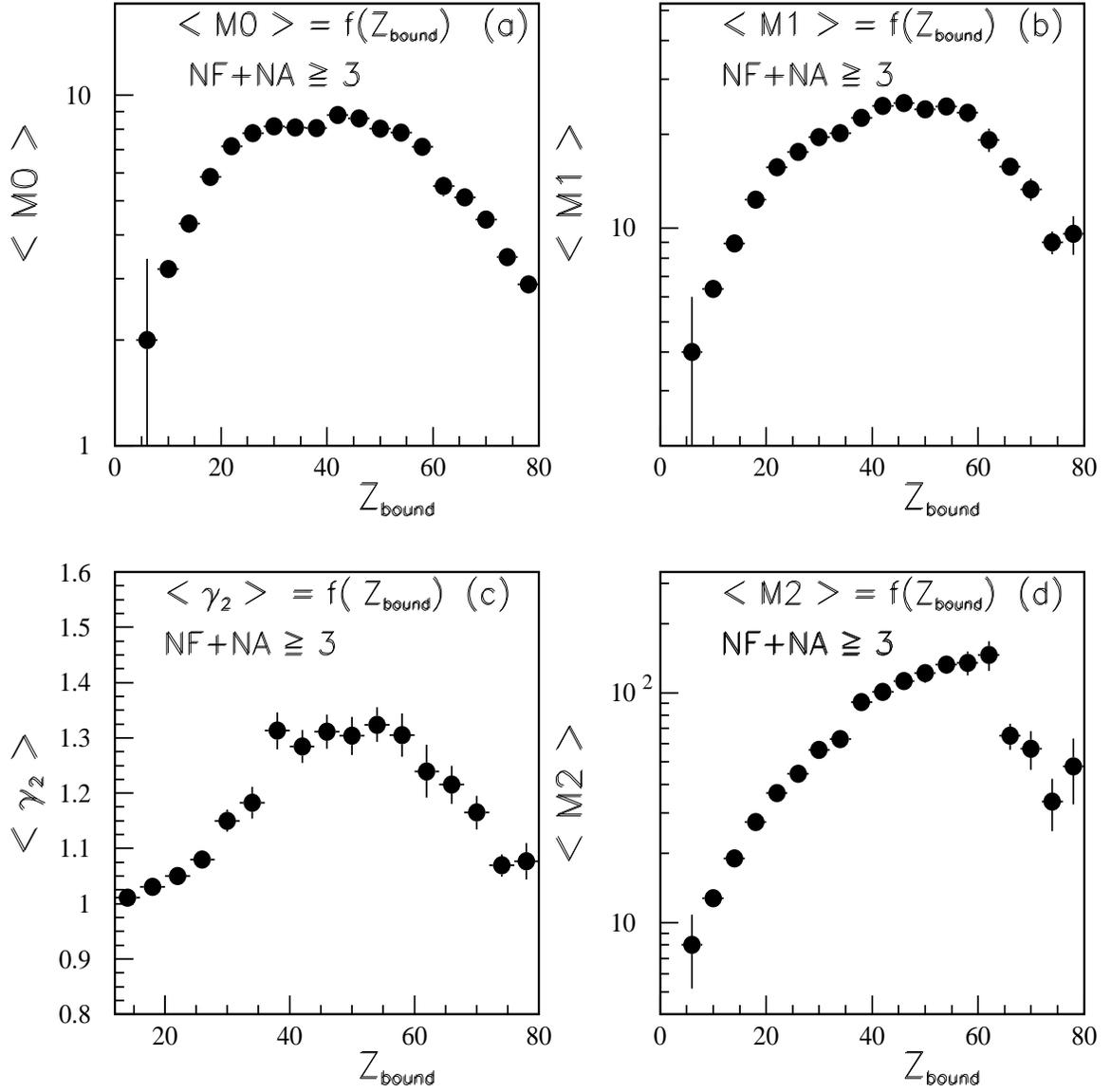}
\caption{Mean values of zeroth moments, $< M_{0} >$ (Fig. 8a), first moments,
$< M_{1} >$ (Fig. 8b) and second moments $< M_2 >$ (Fig. 8d) as a function of
$Z_{bound}$. Part (c) - Variation of the mean values of the conditional
moment $< \gamma_2 >$ with $Z_{bound}$. For the definitions see section 4.3}
\end{figure}

\newpage
\begin{figure}[H]
\vspace{1.cm}
\hspace{1.5cm}
\psfig{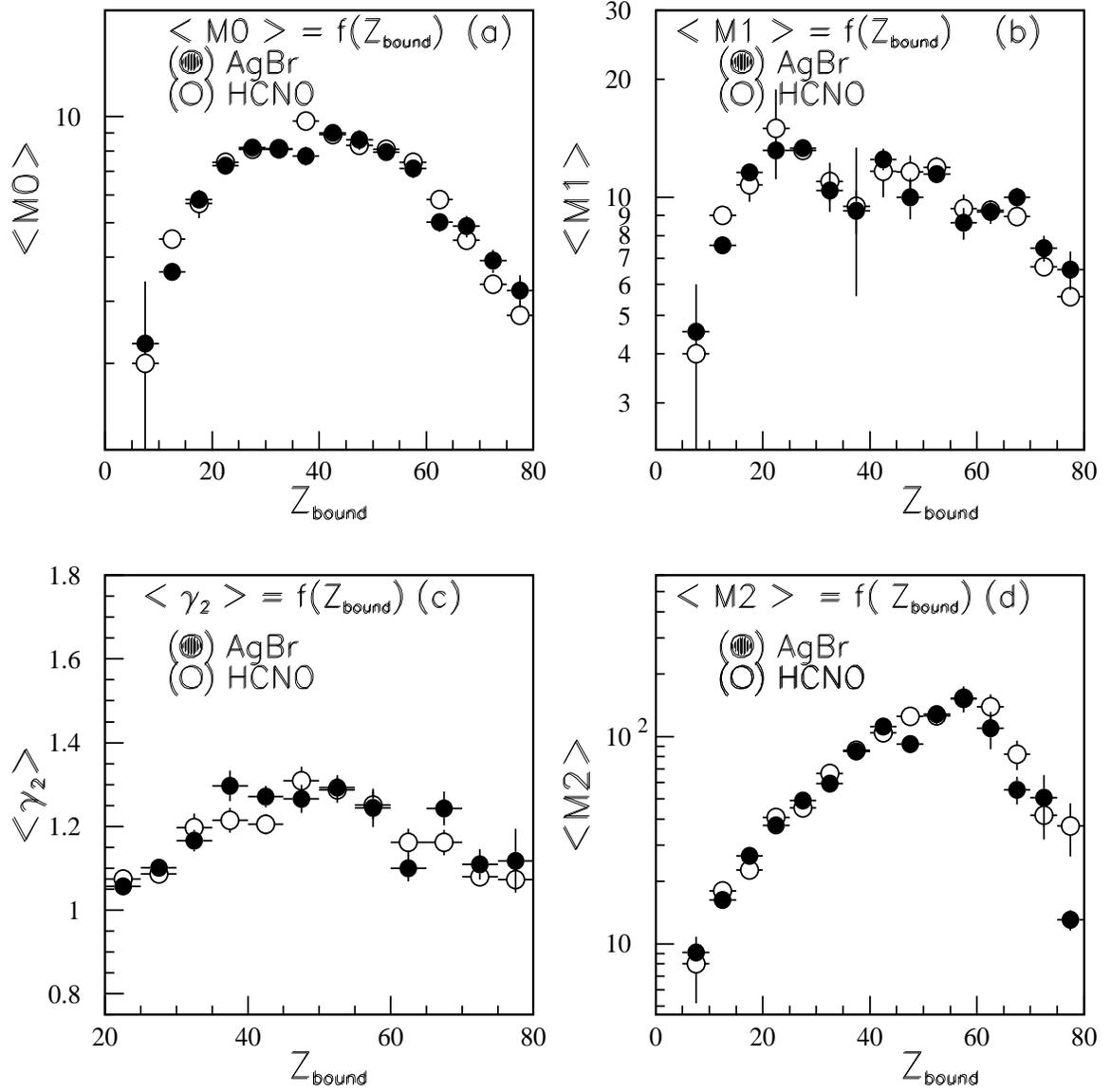}
\caption{Mean values of zeroth moments, $< M_0 >$ (Fig. 9a), first moments,
$< M_1 >$ (Fig. 9b) and second moments $< M_2 >$ (Fig. 9d) as a function of
$Z_{bound}$. Part (c)- Variation of the mean values of the conditional
moment $< \gamma_2 >$ with $Z_{bound}$. The experimental data points have
the same meanings as in Figure 4.
For the definitions of the symbols see section 4.3}
\end{figure}

\newpage

\begin{figure}[H]
\vspace{1.cm}
\hspace{1.5cm}
\psfig{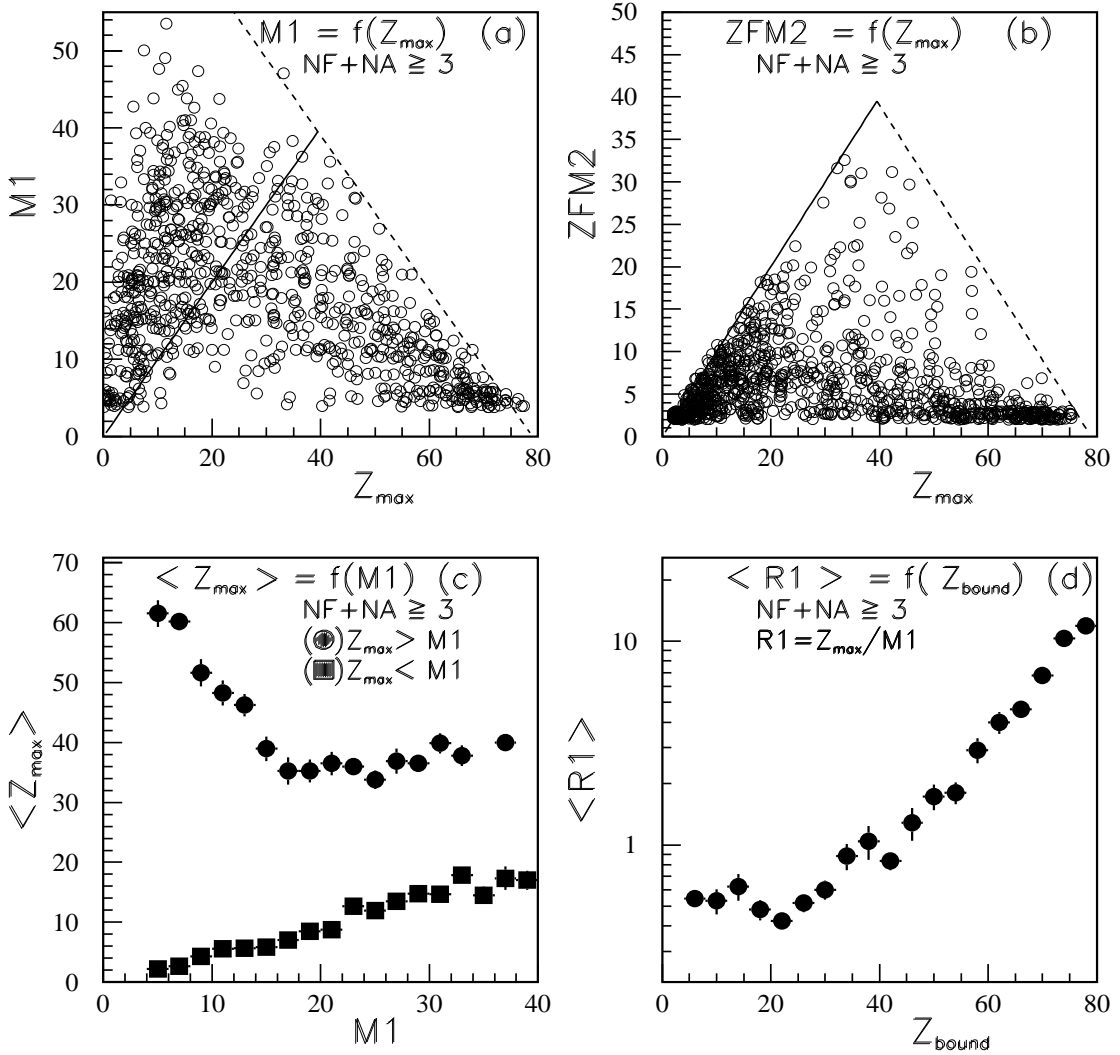}
\caption{ Part (a) - Correlation
between charge on heaviest fragment $Z_{max}$ 
and remaining bound charge $Z_r \equiv M_1$. 
Right diagonal dashed line show charge limit, left full line
shows where $M_1 = Z_{max}$. Part (b) - 
Correlation between heaviest, $Z_{max}$, and second 
heaviest fragments. Diagonal dashed lines show
charge limits, left full line shows where $Z_{FM2}\, =\, Z_{max}$. 
Part (c) - Mean of the charge on 
the heaviest fragment $< Z_{max} >$, as a function 
of the first moments, $M_1$ separated into those
events where $Z_{max}$ is greater or less than $M_1$. Part (d) - Mean
values of the leading fragment ratio 
$< R_1 >\, =\, Z_{max}/M_1$ as a function of
bound charge $Z_{bound}$.}
\end{figure}

\newpage
\begin{figure}[H]
\vspace{1.cm}
\hspace{1.5cm}
\psfig{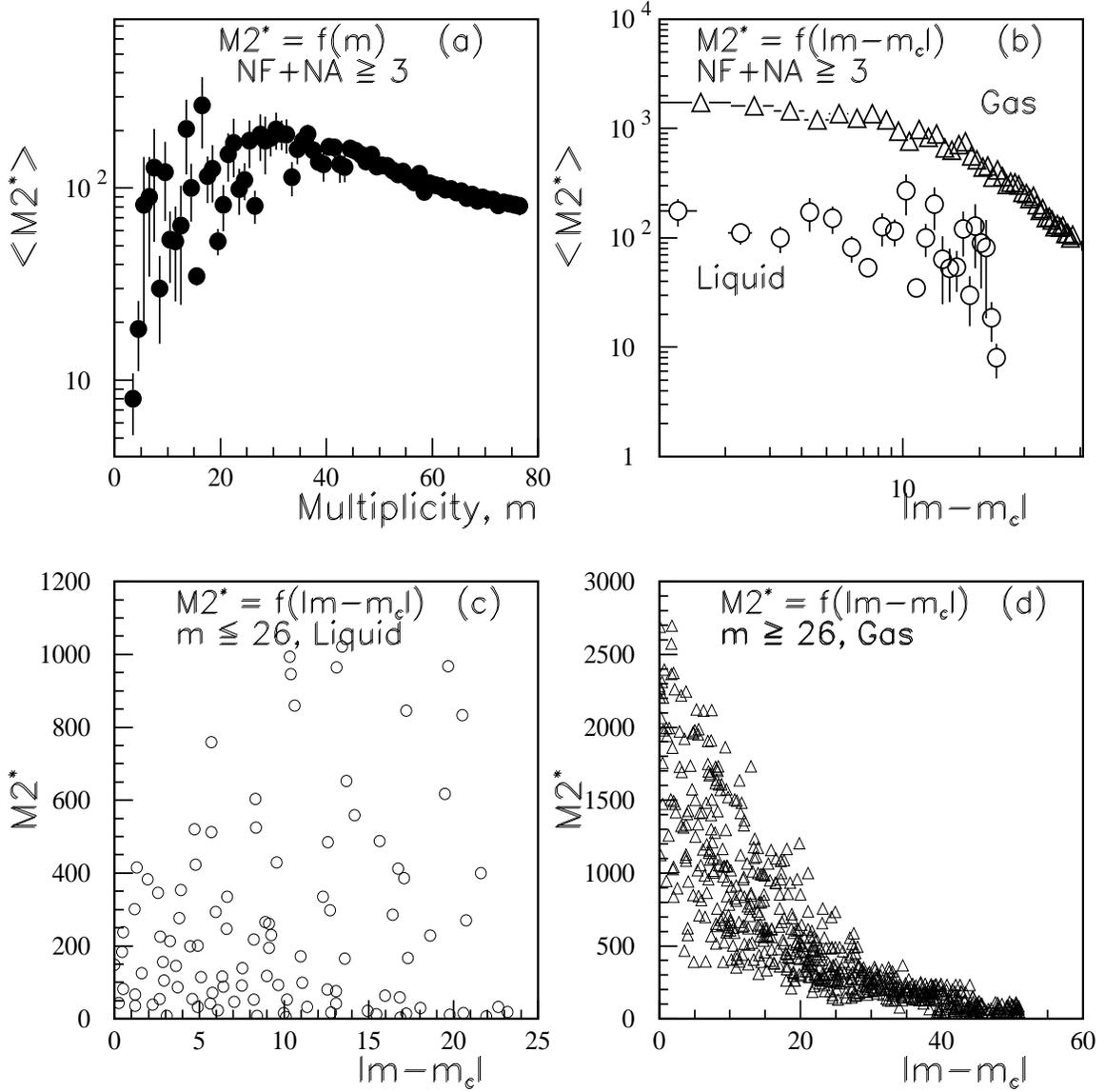}
\caption{Part (a) - Mean second
moments $< M_2^{*} >$ as a function of multiplicity $m$. Part (b) - Mean
second moments $< M_2^{*} >$ as a function of the 
multiplicity difference $\zeta $ assuming $m_c = 26$. 
Part (c) - A scatter plot of second moments $M_2^{*}$ 
as a function of of the multiplicity difference $\zeta $ for liquid
phase. Part (d) - A scatter plot of second moments $M_2^{*}$ as a function of
of the multiplicity difference $|\zeta| $ for gas phase.}
\end{figure}

\begin{figure}[H]
%\vspace{1.cm}
\hspace{1.5cm}
\psfig{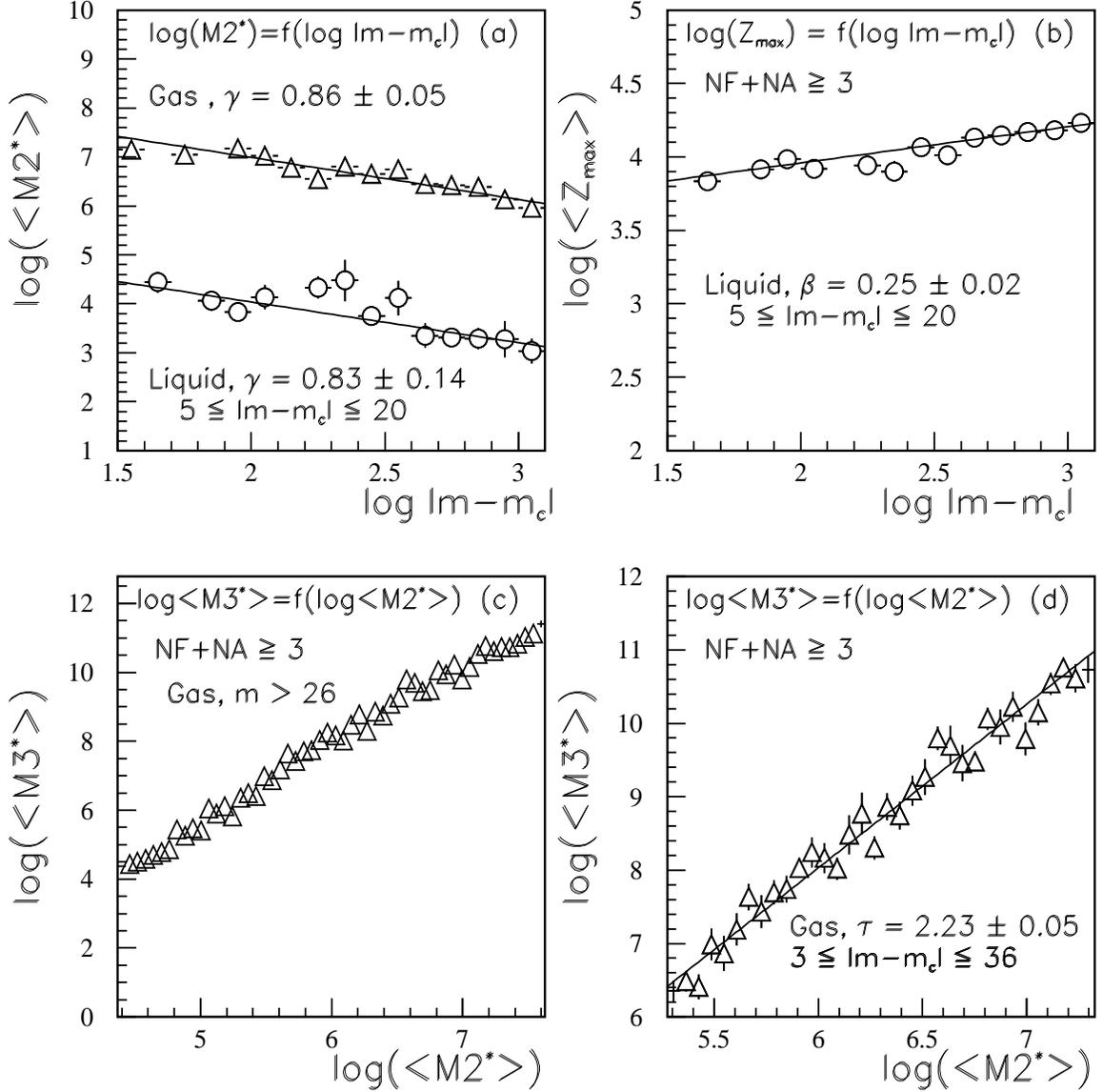}
\caption{Part (a) - A log-log
plot of mean second moments $< M_2^{*} >$ as a function 
of multiplicity difference 
$\zeta$ assuming $m_c = 26$, over a limited range of 
$\zeta$. Linear fits
are shown for both the gas and liquid phases to define 
the exponent $\gamma$ (see Eq. \protect{\ref{eq:m2star}}).
 Part (b) - A log-log plot of the mean values of heaviest
charge, $< Z_{max} >$ , as a function of multiplicity difference $\zeta $,
in the liquid phase, assuming $m_c = 26$. 
A linear fit is shown to define the
exponent $\beta$, Eq. \protect{\ref{eq:zatbeta}}. 
Part (c) - Correlation between mean second and
third moments $< M_2^{*} >$ and $< M_3^{*} >$, for the gas phase, 
assuming $m_c = 26$. 
Part (d) - A log - log plot for correlation between mean second and
third moments $< M_2^{*} >$ and $< M_3^{*} >$, 
for the gas phase, assuming $m_c = 26$ over a limited 
range of $\zeta$. A linear fit is shown to define
the exponent $\tau$, Eq. \protect{\ref{eq:nzattau}}.}
\end{figure}

\end{document}